# TESLA*HERA Based γp and γA Colliders[1]

A. K. Ciftci[a], S. Sultansoy[b,c,d], O. Yavas[a]

[a]*Department of Physics, Fac. of Sciences, Univ. of Ankara, 06100 Ankara, TURKEY*

[b]*Deutsches Elektronen-Synchrotron DESY, Notke Str. 85, D-22607 Hamburg, GERMANY*

[c]*Physics Dept., Faculty of Arts and Sciences, Gazi Univ., 06500 Ankara, TURKEY*

[d]*Institue of Physics, Academy of Sciences, H. Cavid avenue 143, Baku, AZERBAIJAN*

**Abstract**

Main parameters and physics search potential of γp and γA colliders, which will be available due to constructing the TESLA linear electron-positron collider tangentially to the HERA proton ring, are discussed.

*PACS:* 13.61.-r; 24.85.+p; 29.17.+w; 29.20.Dh

*Keywords:* HERA; TESLA; Lepton; Photon; Hadron; Nucleus

## 1. Introduction

It is known that TESLA collider will be a powerful tool for exploration the multi-hundred GeV scale [1]. Different options of this machine, namely, $e^+e^-$, γγ, γe and $e^-e^-$ are complementary to each other and will add an essential new information to that obtained from the LHC. Taking into account the possible polarized proton [2] and nucleus [3] options for HERA itself, the construction of TESLA tangentially to HERA, will provide a number of additional opportunities to investigate lepton-hadron and photon-hadron interactions at TeV scale.

TESLA*HERA based ep collider was proposed ten years ago by B. Wiik and collaborators [4, 5]. In [6] the method to overcome the bunch length limitations on $\beta_p^*$ is developed. This method allow to achieve $L_{ep}=10^{31}$cm$^{-2}$s$^{-1}$ within moderate upgrade of TESLA and HERA parameters [7]. The γp, eA and γA options of the TESLA*HERA complex were considered in the joint DESY-AU study [8]. The luminosity of γp collider has the same order as ep option and estimations show that $L \cdot A=10^{30}$cm$^{-2}$s$^{-1}$ can be achieved at least for light nuclei. In addition to mentioned options, colliding of the TESLA FEL beam with nucleus bunches from HERA will give an interesting possibility to investigate "traditional" nuclear phenomena [9].

---

[1] Presented by S. Sultansoy at the International Workshop on High Energy Photon Colliders

14-17 June 2000, DESY, Hamburg, Germany

Finally, scattering of the high-energy photon beam on a polarised nuclear target can be used for an investigation of the spin content of nucleon [10].

There are a number of reasons favouring a superconducting linac (TESLA) as a source of e-beam for linac-ring type collider [4]. First of all spacing between bunches in warm linacs, which is of the order of ns doesn't match with the bunch spacing in the HERA. Also, the pulse length is much shorter than the ring circumference. Additionally, in the case of TESLA, which use standing wave cavities, one can use both shoulders in order to double electron beam energy, whereas in the case of conventional linear colliders one can use only half of the machine, because the travelling wave structures can accelerate only in one direction

Recently, different options of the TESLA*HERA complex are investigated by the THERA study group [11].

## 2. Main parameters of γp collider

Earlier, the idea of using high energy photon beams obtained by Compton backscattering of laser light off a beam of high energy electrons was considered for γe and γγ colliders (see [12] and references therein). Then the same method was proposed for constructing γp colliders on the base of linac-ring type *ep* machines and rough estimations of the main parameters of γp collisions are given in [13]. In general, from technical point of view γp and γA colliders are similar to TESLA based γe collider. The well-known formulae for γe option [14, 15] are modified for γp collider in [16], where the dependence of different parameters on the distance *z* between conversion region and collision point was analysed and some design problems were considered.

Upgraded parameters of TESLA electron beam and HERA proton beam are given in Tables 1 and 2, respectively. Using these parameters one can easily obtain the center-of-mass energies and luminosity values at z=0 m which are presented in Table 3. Here we take into account the electron-to-photon conversion coefficient 0.65 [12] and a factor of two coming from smallness of the photon bunch transverse sizes.

Using the formulae from [16], below we illustrate main features of γp collisions by considering $E_e$=300 GeV option. As one can see from Figure 1, the luminosity slowly decreases with the increasing *z* (factor ~1/2 at *z*=5 *m*) and opposite helicity values for laser and electron beams are advantageous. Additionally, a better monochromatization of high-energy photons seen by proton bunch can be achieved by increasing the distance *z* (see Fig. 2). Luminosity distribution as a function of γp invariant mass at z=5 m is plotted in Figure 3 for three different choices of laser and electron polarizations. Mean helicity of high energy photons as a function of γp invariant mass is plotted in Figure 4.

In general, there are two possible choices for γp interaction region:
- Head-on collisions,
- Collisions with a small angle (crab crossing).

Also, for both possibilities, there are two options:
- Deflection of the electron beam,
- Without deflection.

For the first option the synchrotron radiation on proton beam line and detector can lead to some problems. In the second option residual electron-proton beam-beam tune shift effect should be kept under control. The work on these subjects is going on. Let us mention that the larger distance between the conversion region and the interaction point has advantages for solving design problems. In difference from the γe option of TESLA, where this distance should be of order of mm, for γp option even at z=5 m we loose only the half of the luminosity (see Fig. 1).

## 3. Main parameters of γA collider

In this option sufficiently high luminosity can be achieved at least for light nuclei because the main limitation comes from intra-beam scattering (IBS) and it is known that emittance growth time is proportional to $Z^2/A$. Also, the scheme with deflection of electron beam after conversion is preferable because it will give opportunity to avoid limitations from $\Delta Q_A$ especially for medium and heavy nuclei. For illustration we consider Carbon nucleus beam with parameters given in last column of the Table II. The dependence of luminosity on the distance between conversion region and interaction point is plotted in Figure 5 for $E_e$=300 GeV. Let us remind that an upgrade of the luminosity by a factor 3-4 may be possible by applying a "dynamic" focusing scheme [6]. Further increasing of luminosity will require the cooling of nucleus beam in the HERA ring [17].

## 4. Physics at γp option

Physics search potential of γp colliders had been discussed in a number of papers:
- UNK*VLEPP [18, 19]
- HERA*LC [20].

More general consideration, including ep, γp, eA and γA options of the TESLA*HERA and Linac*LHC complexes can be found in [8, 21].

Partial list of physics goals of the THERA based γp collider contains:
- Total cross-section at TeV scale, which can be extrapolated from existing low energy data as σ(γp→hadrons) ≈ 100÷200 μb
- Two-jet events, $10^4$ events per working year with $p_t$>100 GeV
- Heavy quark pairs, $10^7 \div 10^8$ ($10^6 \div 10^7$, $10^2 \div 10^3$) events per working year for cc (bb, tt) pair production
- Hadronic structure of the photon
- Single W production, $10^4 \div 10^5$ events per working year
- Excited quarks ($u^*$ and $d^*$) with m≤1 TeV
- Single leptoquarks with m≤0.5 TeV
- Associate wino-squark and gluino-squark production if the sum of their masses is less than 0.5 TeV.

The polarization of converted photons will provide important advantages. In addition, γp collider with longitudinally polarized proton beam will be the powerful tool for investigation of spin content of the proton.

Let us finish this section with the comment on high energy isolated lepton events [22, 23] observed by H1 Collaboration. These events can be interpreted as the result of single t-quark production due to anomalous γ-c-t or γ-u-t interaction [24, 25]. In this case, t-quarks will be copiously produced at γp colliders via γc (sea c-quark) or γu (valence u-quark) fusion:

$$\gamma + c \rightarrow t \rightarrow W^+ + b \rightarrow l^+ + \nu + b,$$
$$\gamma + \bar{c} \rightarrow \bar{t} \rightarrow W^- + \bar{b} \rightarrow l^- + \bar{\nu} + \bar{b}$$

or

$$\gamma + u \rightarrow t \rightarrow W^+ + b \rightarrow l^+ + \nu + b.$$

The former possibility leads to isolated leptons with both signs, whereas in the last case only positive charged leptons are expected. If the alternative, stop scenario [26, 27] is valid, the large number of events with an isolated positive (negative) charge leptons and large missing transverse momentum will be produced at $e^+p$ ($e^-p$) option of the TESLA*HERA complex. Therefore, γp and ep options are complementary to each other.

**5. Physics at γA option**

Preliminary list of physics goals contains:
- Total cross-section to clarify real mechanism of very high energy γ-nucleus interactions
- Investigation of a hadronic structure of the proton in nuclear medium
- According to the VMD, proposed machine will also ρ-nucleus collider
- Formation of quark-gluon plasma at very high temperatures but relatively low nuclear density
- Gluon distribution at extremely small $x_g$ in nuclear medium (γA→QQ+X)
- Investigation of both heavy quark and nuclear medium properties (γA→J/Ψ(Y)+X, J/Ψ(Y)→$l^+l^-$
- Existence of multi-quark clusters in nuclear medium and a few-nucleon correlation.

In our opinion γA collider is the most promising option of the TESLA⊗HERA complex, because it will give unique opportunity to investigate small $x_g$ region in nuclear medium [28]. Indeed, due to the advantage of the real γ spectrum heavy quarks will be produced via γg fusion at characteristic

$$x_g \approx \frac{5 \times m_{c(b)}^2}{0.8 \times (Z/A) \times s_{ep}},$$

which is approximately $(2 \div 3) \cdot 10^{-5}$ for charmed and $(2 \div 3) \cdot 10^{-4}$ for beauty hadrons. The number of cc and bb pairs, which will be produced in γC collisions, can be estimated as $10^6 \div 10^7$ and $10^5 \div 10^6$ per working year, respectively. Therefore, one will be able to investigate small $x_g$ region in details. For this reason very forward detector in γ-beam direction will be useful for investigation of small $x_g$ region due to registration of charmed and beauty hadrons.

**6. Conclusion**

Advantages of the THERA based γp and γA colliders are following:
i. Better spectrum of converted photons versus Weizsacker-Williams (quasi-real) photons,
ii. Good kinematics, namely, energy of the γ beam is close to the energy of the valence quark,
iii. High polarization of real photons
vi. These options will be unique to THERA complex and can not be realized at LEP*LHC.

Finally, as it was mentioned by Professor Wiik [29], TESLA*HERA complex will be the first representative of new linac-ring type electron-proton colliders, which seems today to be the sole way to multi-TeV scale in lepton-hadron and photon-hadron collisions (see review [30]).

**Acknowledgements**

We are grateful R. Brinkmann, A. Celikel, I. Ginzburg, U. Katz, M. Klein, M. Leenen, V. Serbo, Y. Sirois, V. Telnov, D. Trines, G.-A. Voss, A. Wagner and F. Willeke for useful discussions and valuable remarks. This work is supported by Turkish State Planning Organisation under the Grant No DPT-97K-120420 and DESY.

**Appendix: Utilization of γ-beam for fixed target experiments**

The advantage in energy spectrum and polarization of Compton back-scattered photons open the unique opportunity to measure polarised gluon distributions using polarized nucleus targets [31-34]. This option for TESLA is investigated in [10], where the formation of high energy photon beam specially for this purpose was considered. In our opinion, it is a matter of interest to reconsider the subject taking into mind the γ-beams, which will be produced for γγ and γe modes of the TESLA or γp and γA modes of the THERA. Main modification is caused by number of converted photons, which becomes $\approx 10^{10}$ instead of $\approx 10^4$ considered in [10]. Therefore, instead of the deuterated butanol target with the density $T_n \approx 10^{25}$ one should use the thick target with $T_n \approx 10^{19}$. The integrated luminosity of γN collisions will be of order of 1 fb$^{-1}$ per working year and polarization asymmetry can be measured with statistical precision $\approx 1\%$ (0.3%) using $D^*$-tagging (single muon tagging). Of course, detailed analysis of design problems (γ-beam transportation, target and detector issues etc) is needed.

Table 1
Upgraded TESLA Parameters

| Electron Energy, GeV | 300 | 500 | 800 |
|---|---|---|---|
| Number of e per bunch, $10^{10}$ | 1.8 | 1.5 | 1.41 |
| Number of bunches/pulse | 4640 | 4950 | 4430 |
| Beam power, MW | 40 | 30 | 40 |
| Bunch length, mm | 1 | 0.4 | 0.3 |
| Bunch spacing, ns | 192 | 192 | 192 |
| Repetition rate, Hz | 10 | 5 | 5 |

Table 2
Upgraded Parameters of Proton Carbon Beams

| | p | C |
|---|---|---|
| Beam energy, TeV | 0.92 | 5.52 |
| No. of particles/bunch, $10^{10}$ | 30 | 0.8 |
| No. of bunches/ring | 90 | 90 |
| Bunch separation, ns | 192 | 192 |
| $\varepsilon_p^N$, $10^{-6}$ $\pi$·rad·m | 0.8 | 1.4 |
| $\beta_p^*$, cm | 20 | 20 |
| $\sigma_{x,y}$ at IP, $\mu m$ | 12.5 | 23 |
| $\sigma_z$, cm | 15 | 15 |

Table 3
Maximal center-of-mass energy and luminosity for various THERA $\gamma p$ collider options

| $E_e$, GeV | 300 | 500 | 800 |
|---|---|---|---|
| $E_\gamma^{max}$, GeV | 250 | 415 | 660 |
| $\sqrt{s_{\gamma p}}^{max}$, GeV | 0.96 | 1.23 | 1.56 |
| $L_{\gamma p}$ at z=0, $10^{31} cm^{-2} s^{-1}$ | 1.66 | 0.74 | 0.62 |

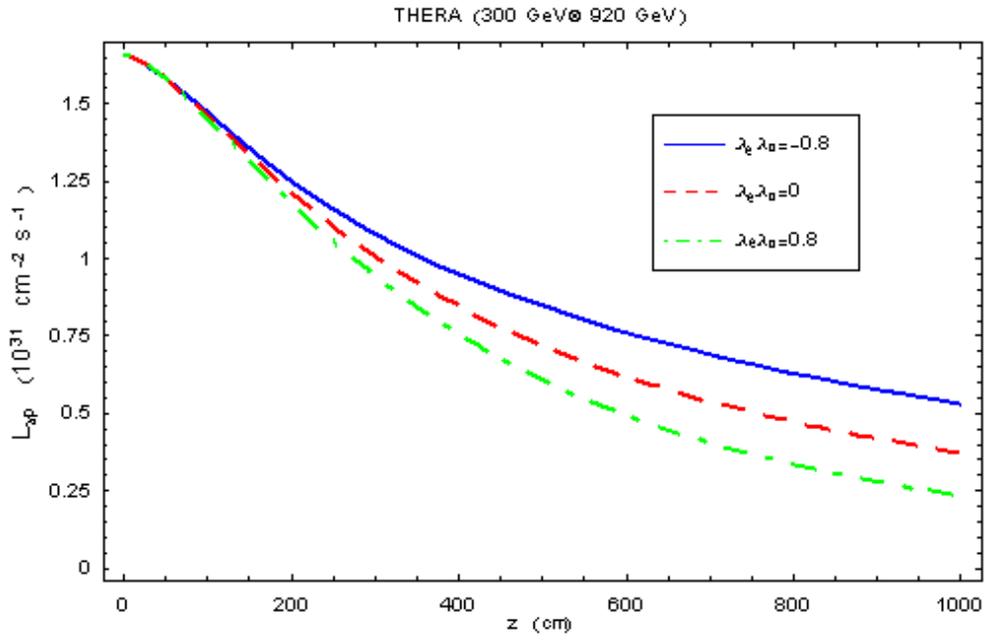

Fig. 1. Dependence of the luminosity on the distance z for γp collider

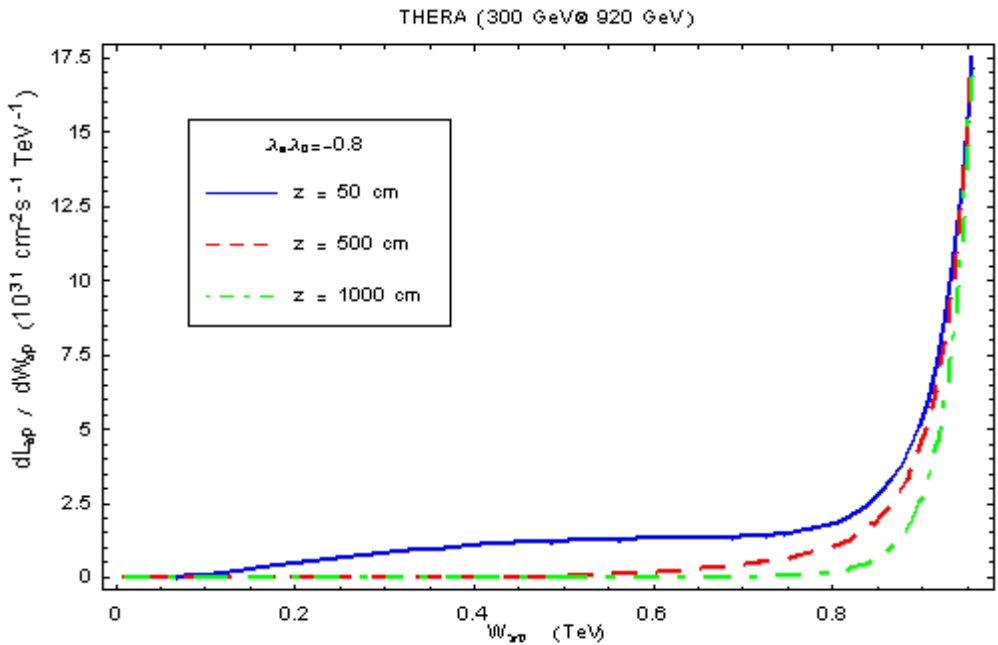

Fig. 2. Luminosity distribution as a function of γp invariant mass for three different values of z

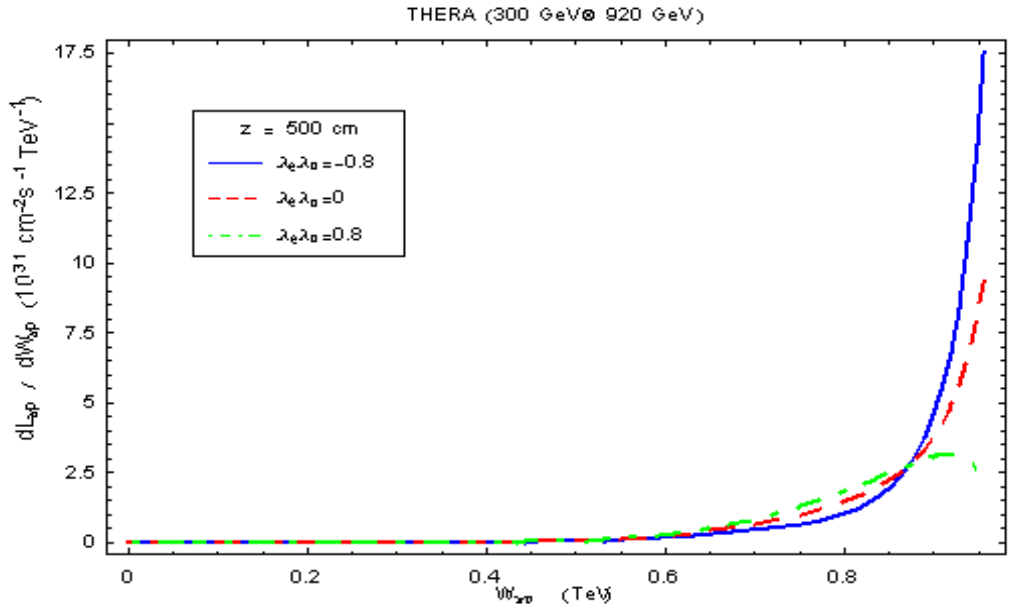

Fig. 3. Luminosity distribution as a function of γp invariant mass at z=5m for choice of three different electron polarization

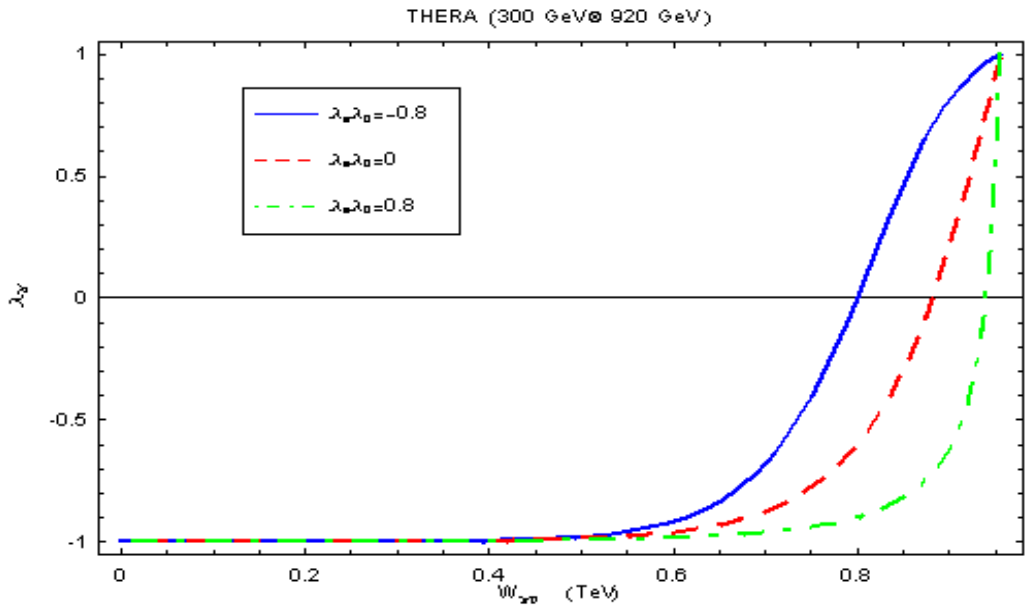

Fig. 4. High energy photon helicity as a function of γp invariant mass

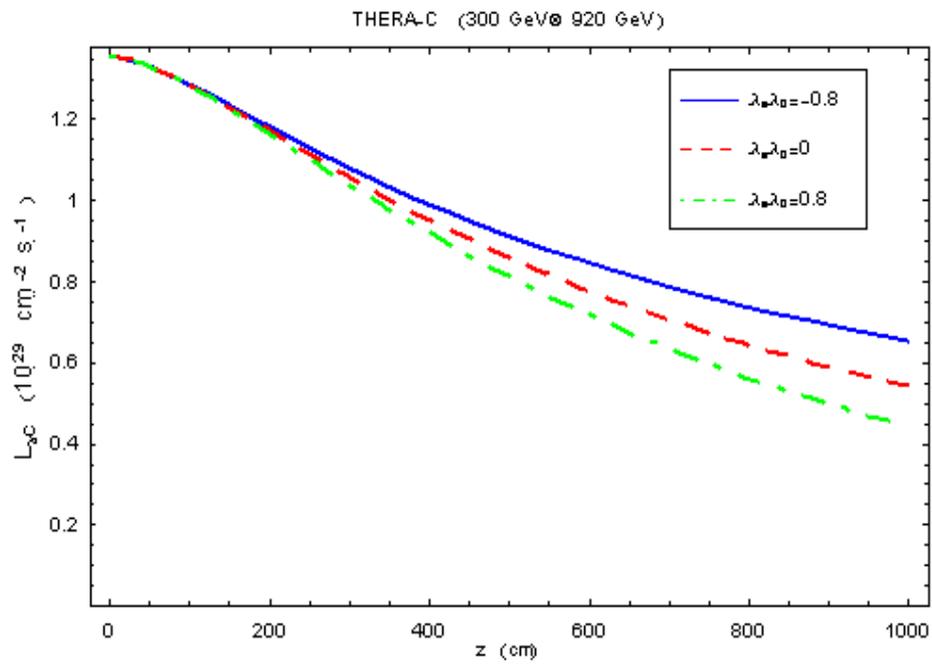

Fig. 5. The dependence of luminosity on the distance z for γC collider